\documentclass[12pt]{iopart}

\usepackage{iopams}  
\usepackage{cite}
\usepackage{graphicx}

\begin{document}

\title[Augmented reality for visualising magnetic field topology in fusion plasmas]
{Augmented reality system for visualising magnetic field topology and charged-particle trajectories in magnetic fusion plasmas}

\author{Akinobu Matsuyama$^{1}$}

\address{$^{1}$Graduate School of Energy Science, Kyoto University, Gokasho, Uji 611-0011, Japan}
\ead{akinobu.matsuyama@gmail.com}

\begin{abstract}
A cost-effective augmented reality (AR) system is presented for visualising three-dimensional magnetic field structures and charged-particle trajectories in magnetically confined fusion plasmas. The system presented in this study integrates an orbit-following simulation code with a marker-based AR framework using a web camera and the OpenCV library. By synchronizing the time step of the simulation with the frame rate of the camera, the trajectories are continuously updated and superimposed in real time onto the camera image. Through the interactive operation of manipulating the web camera, users can observe three-dimensional structures, such as magnetic islands, from various positions and viewing angles. Such an AR environment supports an interactive means of exploring three-dimensional spatial structures that can be difficult to interpret from two-dimensional representations alone. It also provides a common visual representation that can be shared through a display by researchers and students with diverse backgrounds in physics, engineering, and related fields. The developed system has been used in practical exercises at the JT-60SA International Fusion School, with exploratory feedback from students.
\end{abstract}

\submitto{European Journal of Physics}
\maketitle

\section{Introduction}
Fusion energy has attracted attention as a candidate energy source available to future generations, and research activities including the ITER project \cite{Barabaschi2026} are being carried out. A magnetic fusion reactor is a plant that combines various technologies, such as plasma physics, superconducting magnets, material science, and nuclear engineering. To support long-term scientific and technological development, human resource development and outreach activities are being actively conducted at universities, graduate schools, and research institutes.

In magnetic confinement fusion experiments, high-temperature plasmas of hydrogen isotopes exceeding $10^8$ K are confined by magnetic fields. The basic principle of confinement is the formation of nested toroidal surfaces by magnetic field lines. Because the magnetic field is divergence-free, magnetic field lines, which are the streamlines of the magnetic field, have neither starting points nor ending points. It can be shown that if one attempts to keep their trajectories inside a finite volume, the equilibrium configuration of the plasma necessarily takes a toroidal configuration \cite{Kruskal1958}. In a simple circular toroidal magnetic field, ions and electrons undergo magnetic drifts in opposite directions, which causes charge separation and prevents the formation of an equilibrium configuration capable of confining the plasma \cite{Hazeltine1992}. Therefore, the magnetic field lines are twisted helically to form toroidal surfaces, called magnetic surfaces. The ratio between the number of rotations made by a field line around the minor axis and that around the major axis of the torus is called the rotational transform. A tokamak configuration \cite{Wesson} generates the rotational transform by a current flowing in the plasma itself, whereas a stellarator configuration \cite{Wakatani} provides the rotational transform by twisting the coil shapes.
  
Tokamak devices possess an approximate symmetry in the direction around the major axis; therefore, it is relatively easy to grasp the structure of the magnetic surfaces from two-dimensional projections. However, three-dimensional structures that break this symmetry, called magnetic islands \cite{Taylor2015}, often arise spontaneously due to magnetohydrodynamic instabilities or are driven externally. In the case of stellarator devices, the situation is more complicated because the equilibrium configuration itself has a three-dimensionally twisted structure. Two-dimensional projections and Poincar\'{e} plots are often used to represent the three-dimensional structures formed by magnetic field lines. However, understanding these representations requires prior knowledge, and the structures may also be misinterpreted as planar figures. 

One way to address this problem is to use three-dimensional (3D) visualisation software. In the field of magnetic confinement fusion, software packages such as ParaView \cite{Paraview} are widely used. These tools make it possible to manipulate the viewpoint and observe cross sections easily, but a certain level of familiarity is still needed for controlling the viewpoint with a mouse and for understanding spatial structures from a screen display. 

The 3D visualisation has also been combined with immersive virtual reality (VR) technologies in magnetic confinement fusion research. Ohtani {\it et al.} \cite{Ohtani2011} developed the VR environment called a CAVE system, in which simulation results, including magnetic field lines and particle trajectories, can interactively be visualised together with the geometry of the Large Helical Device (LHD), a stellarator-type magnetic confinement device in Japan. Recently, Ohno {\it et al.} \cite{Ohno2024} developed a VR visualisation using head-mounted displays, providing an immersive environment for interactive visualisation of LHD plasma data. These approaches provide powerful ways for exploring complex 3D plasma structures, although they require dedicated VR hardware.

In this study, we focus on the tokamak configuration and explore a complementary approach based on augmented reality (AR) \cite{Azuma}, with an emphasis on simple deployment and shared visualisation. We develop a method for interactively visualising magnetic field lines and charged-particle trajectories by combining an orbit-following simulation code with a marker-based AR system using a standard web camera and the OpenCV library \cite{Bradski2000}. Users can change their viewpoint by manipulating the pose of the camera, while the visualisation is presented on a common display that can be observed simultaneously by multiple users. The distinctive feature of the present approach is therefore not three-dimensional visualisation for fusion plasma application itself, but its implementation as a low-cost, marker-based shared-display AR environment. This configuration is particularly suitable for educational implementation, group activities, and outreach, where students with diverse backgrounds in physics, engineering, and related fields can observe and discuss the same visual information.

The remainder of this paper is organized as follows. In section \ref{sec:magnetic}, we describe the tracking of magnetic field lines and charged-particle trajectories in a magnetic confinement fusion device, which is the target of the visualisation in this study, and discuss the associated difficulties. In section \ref{sec:AR}, we introduce the AR system developed in this study. In section \ref{sec:Results}, we discuss the visualisation affordances of the AR system through a comparison with conventional visualisation methods based on two-dimensional projections and Poincar\'{e} plots. In section \ref{sec:DC}, we present the conclusions and discuss the remaining issues.

\section{Magnetic Field Lines and Charged-Particle Trajectories}\label{sec:magnetic}
To sustain fusion reactions using a magnetic confinement fusion device, it is necessary to confine a high-temperature and high-pressure plasma satisfying the magnetohydrodynamic equilibrium condition $\nabla p = {\bf J}\times {\bf B}$ \cite{Grad1967}, where $p$ is the plasma pressure, ${\bf B}$ is the magnetic field, and ${\bf J}$ is the current density. Because no material exists that can withstand the heat flux from the high-temperature plasma of interest in controlled fusion, the first problem in understanding magnetic confinement fusion is to find toroidal equilibrium solutions that can keep a plasma with finite pressure away from the wall and to understand the behaviour of magnetic field lines and charged particles in such equilibria. From the MHD equilibrium equation, one obtains ${\bf B}\cdot \nabla p=0$, implying that the plasma pressure is constant along a magnetic field line. When the magnetic field-line flow is integrable, the field lines lie on nested flux surfaces that can be labelled by a constant of motion $\psi$, which corresponds to the magnetic flux enclosed by the toroidal surface. The plasma pressure can be expressed as a function of only $\psi$, i.e., $p=p(\psi)$. 

\begin{figure}[htbp]
  \begin{center}
    \includegraphics[clip,width=14cm]{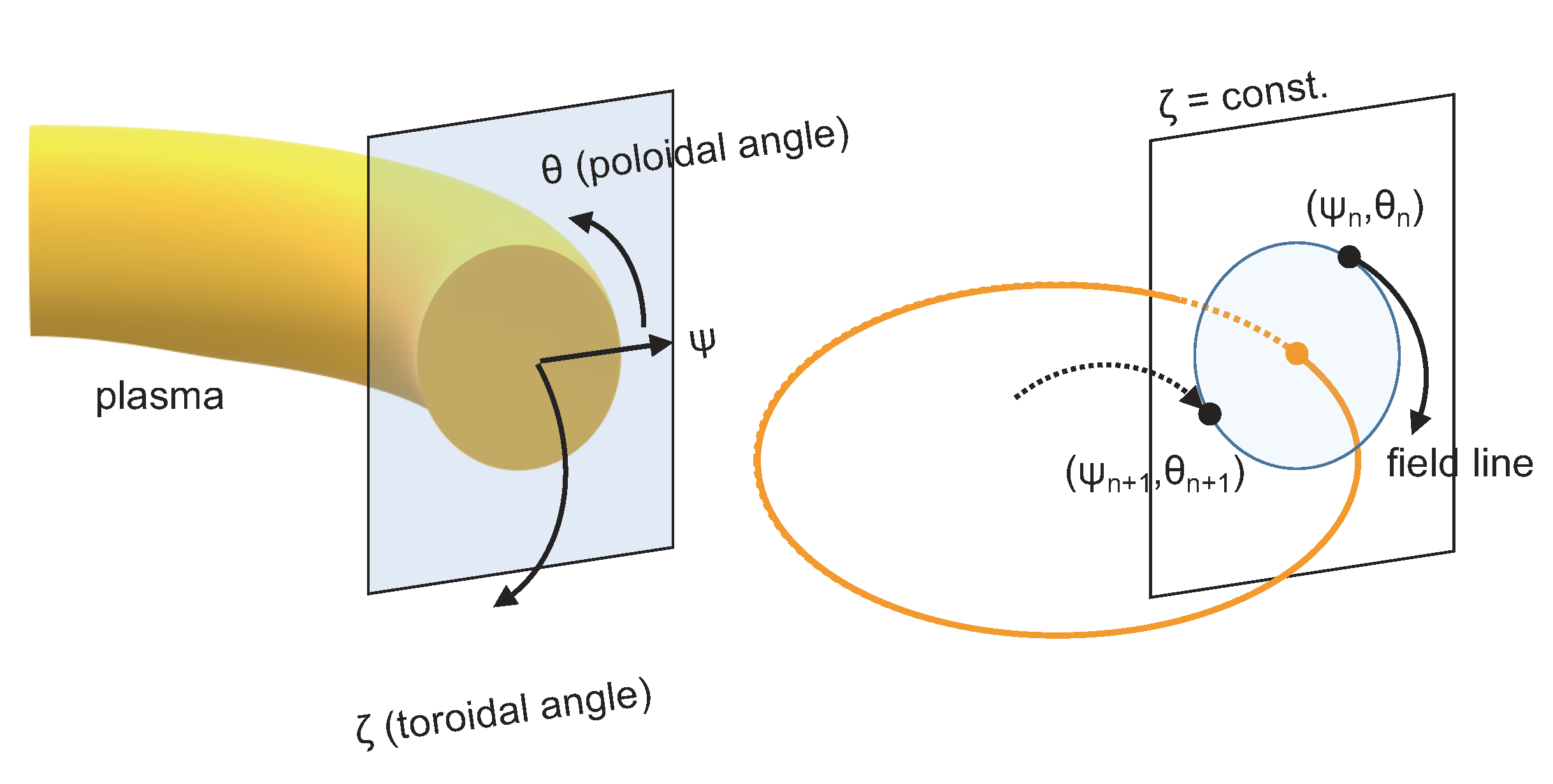}
    \caption{(a) Coordinate system used to describe the toroidal geometry. (b) Schematic illustration of the principle of the Poincar\'{e} plot.}
    \label{fig:Poincare}
  \end{center}
\end{figure}

A general curvilinear coordinate system $(\psi,\theta,\zeta)$, as shown in Fig. \ref{fig:Poincare}(a), is convenient for analysis, where one coordinate is the constant of motion $\psi$ of the field line trajectory and the other two are angular coordinates around the major and minor axes of the torus: toroidal angle $\zeta$ and poloidal angle $\theta$ \cite{Flux}. Here, when the transformation law between the Cartesian and general curvilinear coordinate systems is expressed as ${\bf x}={\bf x}(\psi,\theta,\zeta)$, the transformation satisfies the double periodicity ${\bf x}(\psi,\theta+2\pi,\zeta)={\bf x}(\psi,\theta,\zeta)$ and ${\bf x}(\psi,\theta,\zeta+2\pi)={\bf x}(\psi,\theta,\zeta)$. In systems with such double periodicity, it is convenient to use Poincar\'{e} plots to discuss the behaviour of magnetic field lines. Figure \ref{fig:Poincare}(b) shows the principle of the Poincar\'{e} plot. When a magnetic field line is repeatedly followed with respect to the turns around the major axis of the torus, the position of the field line passing through a plane $\zeta={\rm const.}$ is recorded. Depending on the pitch $\Delta\theta /\Delta \phi$ of the field line, the points rotate in the poloidal $(\theta)$ direction on the mapped plane. If the operation that obtains the intersection point $(\psi_{n+1},\theta_{n+1})$ at the $(n+1)$-th turn from the intersection point $(\psi_n,\theta_n)$ at the $n$-th turn is regarded as a map, the field line trajectory can be abstracted as a two-dimensional area-preserving map, whose dynamical properties can be analysed in a manner similar to that of the standard map \cite{Chirikov1979,Greene1979}.

The charged-particle trajectories in toroidal magnetic geometry are another target of visualisation in this study. The motion of charged particles perpendicular to the magnetic field is constrained to the Larmor gyroradius. Hence high-energy ions in a toroidal magnetic configuration do not immediately collide with the wall. However, when a magnetic field is bent into a torus, inhomogeneity of the magnetic field necessarily appears between the inner and outer sides of the toroidal curvature, and charged particles drift slowly in the cross-field direction. Such a drift is analysed in terms of the motion of the centre of the Larmor gyration, called the {\it guiding centre} \cite{Northrop}. When the Larmor radius is sufficiently small compared with the scale length of the system, the fast gyration itself is often omitted from the treatment, and only the trajectory of the guiding centre is discussed \cite{Cary2009}. The variation in the magnetic-field strength along the guiding-centre trajectory acts as a potential well, and particles with small velocities in the direction of the magnetic field are trapped in this well structure. When the trajectories of such trapped particles are projected onto a plane $\zeta={\rm const.}$, the well-known trajectories of {\it banana particles} are obtained [see Fig. \ref{fig:ITER_AR}(a)].

A large number of numerical simulation codes have been developed and routinely used to analyse magnetic field lines and charged-particle trajectories in magnetic confinement fusion devices. These codes have functions for analysing trajectories by means of two-dimensional projections and Poincar\'{e} plots. Such visualisation methods are sufficiently mature. However, when the problem becomes complicated, grasping the three-dimensional spatial structure and relating it to experimental observations is a task that imposes substantial spatial-processing demands, even on experts. For beginners, understanding the representation of magnetic field and charged-particle trajectories requires a certain amount of prior knowledge. From this point of view, especially as an educational approach, this work introduces a visualisation system using AR for exploring the basic principle of magnetic confinement.

\section{Building the AR environment}\label{sec:AR}
\begin{figure}[htbp]
  \begin{center}
    \includegraphics[clip,width=14.0cm]{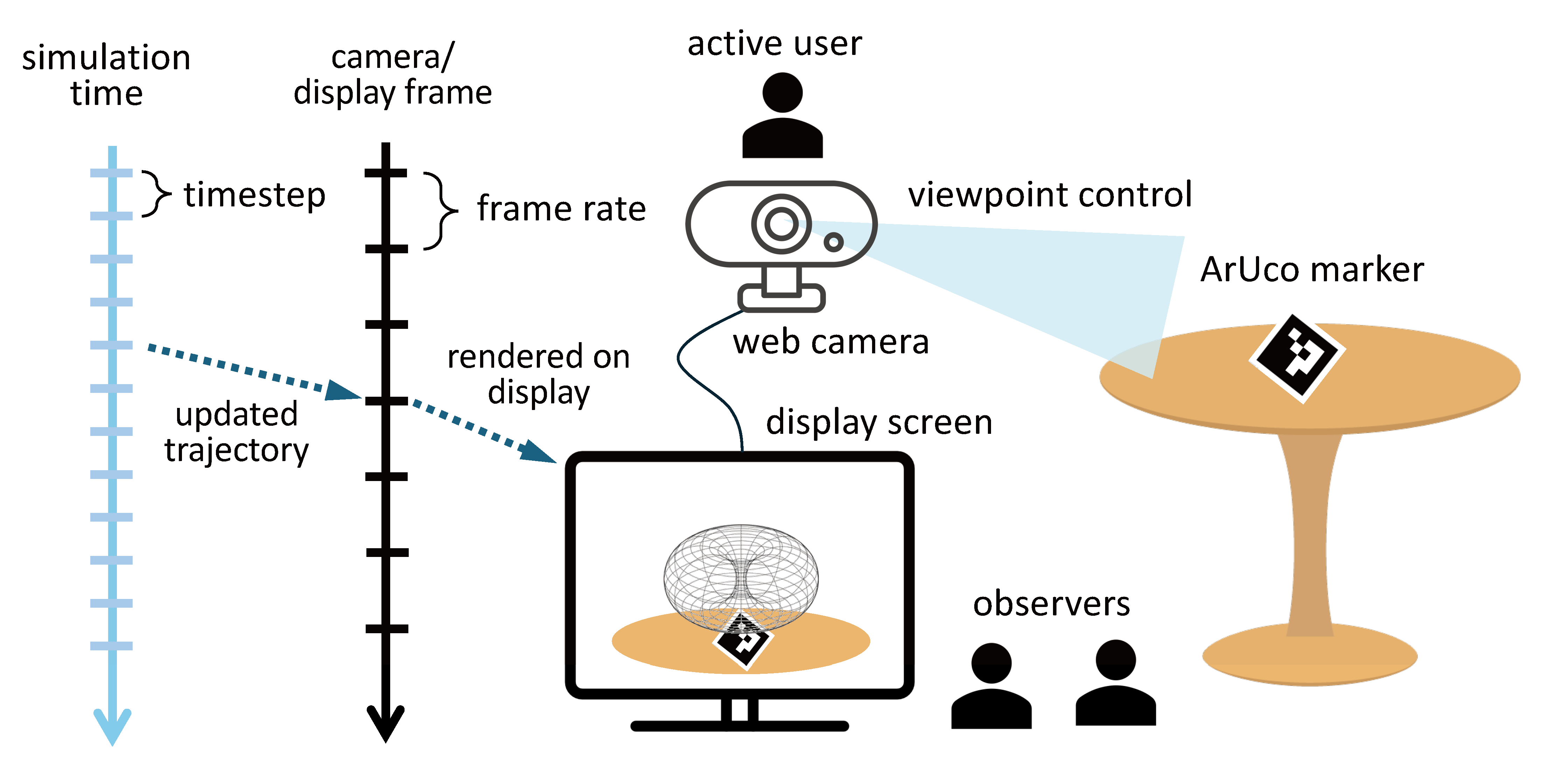}
    \caption{Schematic illustration of the developed AR system.}
    \label{fig:ARsystem}
  \end{center}
\end{figure}

In this section, we describe an AR visualisation system for magnetic field lines and charged-particle trajectories. AR is characterized by three defining features \cite{Azuma}: (i) it combines real and virtual objects, (ii) it is interactive in real time, and (iii) it is registered in three dimensions. Figure \ref{fig:ARsystem} shows a schematic illustration of the AR system developed in this study. In this system, the three-dimensional trajectory information obtained by simulation is superimposed onto images captured by a web camera, thereby satisfying condition (i). Furthermore, instead of drawing a static image, the system synchronizes the simulation time step with the frame rate of the camera and represents the three-dimensional trajectories of the magnetic field lines and charged particles as motion, that is, as temporal evolution. Thus conditions (ii) and (iii) are satisfied. The focus of the present system is to provide a complementary three-dimensional representation of structures conventionally analysed using Poincar\'{e} plots or three-dimensional visualisation software as a real-time experience that includes the bodily experience of the user manipulating the web camera with visual feedback.

The system consists of an orbit-following simulation code for a magnetic confinement fusion device, an AR processing module using the OpenCV library, a web camera, and a display. We employed the guiding-centre orbit-following code ETC-Rel \cite{Matsuyama2014} as the simulation code for magnetic field lines and particle orbits. For visualisation, a Logicool C1000eR web camera was used, providing a horizontal resolution of 4096 pixels and vertical resolution of 2160 pixels at the frame rate of 30.0 fps. The calculations and AR visualisation presented in this study were performed on a MacBook Air equipped with an Apple M2 processor and 24 GB of memory, running macOS Sonoma 14.8.9, Python 3.12.14, and OpenCV 5.0.0. The AR visualisation is based on a marker-based approach using fiducial markers, specifically ArUco markers \cite{Garrido2014}. The OpenCV library \cite{Bradski2000} was used for camera calibration, marker detection, and pose estimation. A marker from the predefined ArUco {\tt DICT$\_$4X4$\_$50} dictionary was used for pose estimation. The camera was calibrated in advance using the standard OpenCV camera-calibration procedure with a $9\times 6$ chessboard pattern with a square size of 28 mm printed on A4 paper. The resulting intrinsic camera matrix and lens-distortion coefficients were used for the subsequent pose estimation and projection. The system requires only a standard web camera and printed fiducial markers, enabling straightforward deployment in laboratory and educational environments without the need for specialized hardware.

An ArUco marker is printed on a planar surface such that all four corners are fully visible within the camera field of view. The plane defined by the marker is used to establish a reference coordinate system, which we define as the world coordinate system. In this coordinate system, the marker plane is assumed to lie on the constant $Z$-plane. Let ${\bf X}_i^{(w)}=(X_i,Y_i,Z_i)^\top$ $(i=1,2,3,4)$ denote the 3D coordinates of marker corners in the world coordinate system, and ${\bf x}_i=(u_i,v_i)^\top$ their corresponding 2D image coordinates. The camera pose is obtained by solving the PnP problem \cite{Marchand2016}:
\begin{equation}
 {\bf x}_i \sim {\sf K} \cdot \left( {\sf R} \cdot {\bf X}_i^{(w)} + {\bf t} \right),
\end{equation}
where $({\bf R},{\bf t})$ represent the rotation and translation from the world coordinate system to the camera coordinate system, respectively. The intrinsic camera matrix  ${\sf K} \in {\mathbb R}^{3\times 3}$ is defined as follows:
 \begin{equation}
{\sf K} = \left( \begin{array}{ccc} f_x & 0 & c_x \\ 0 & f_y & c_y \\ 0 & 0 & 1 \\ \end{array} \right), 
\end{equation}
where $f_x$ and $f_y$ are the focal lengths in pixel units, and $(c_x, c_y)$ denotes the principal point. The rotation vector obtained from the PnP solution is converted into a rotation matrix using the Rodrigues formula \cite{Forsyth}. 

The magnetic field lines and charged-particle trajectories computed by the ETC-Rel code are represented in the simulation coordinate system, which corresponds to the physical scale of the magnetic fusion device. Let ${\bf P}_j^{(s)} =(x_j, y_j,z_j)^\top$ denote a point on the particle trajectory in the simulation coordinates. To embed the simulation results into the AR scene, a scaling transformation is applied to map the simulation coordinates to the world coordinate system: ${\bf P}_j^{(w)}=s{\bf P}_j^{(s)}$. The scaling factor $s$ determines the displayed size of the simulated device and trajectories relative to the printed marker. In practice, $s$ was adjusted so that the device geometry could be viewed conveniently within the field of view when using the A4-sized printed marker. Because the same scaling factor is applied uniformly to all three spatial coordinates, changing $s$ changes the overall displayed size but does not distort the relative three-dimensional geometry. The transformed points are then mapped into the camera coordinate system as ${\bf P}_j^{(c)}={\sf R}\cdot {\bf P}_j^{(w)} + {\bf t}$. Finally, the 3D points in camera coordinates are projected onto the 2D image plane using the pinhole camera model \cite{Forsyth}. The AR processing module was implemented in Python and interfaced with ETC-Rel through three-dimensional trajectory coordinates $(x,y,z)$, which are transferred at each simulation step. The AR visualisation procedure itself does not depend on the internal formulation of ETC-Rel and can therefore, in principle, be coupled to other simulation codes that provide equivalent three-dimensional trajectory data.

In this study, the history of the entire particle trajectory is visualised as a polyline composed of projected 2D points. Each segment is rendered only if it satisfies the visibility condition described later. The current particle position is highlighted as a filled circle. The processing is performed in a frame-by-frame loop synchronized with the camera acquisition rate. For each frame, the simulation advances by one time step, and the corresponding trajectory is updated and rendered. As shown in Fig. \ref{fig:ARsystem}, the relationship between simulation time and real time is controlled by adjusting the simulation time step relative to the camera frame rate, allowing the temporal evolution to be displayed continuously. The output is displayed on a monitor connected to the system. By physically manipulating the camera, the user can interactively change the viewpoint and observe the particle trajectories and magnetic field structures in real time.

\section{Visualisation Results}\label{sec:Results}
In this section, we describe the visualisation results of the magnetic field lines and charged-particle trajectories using the developed AR system.

\begin{figure}[htbp]
  \begin{center}
    \includegraphics[clip,width=14.0cm]{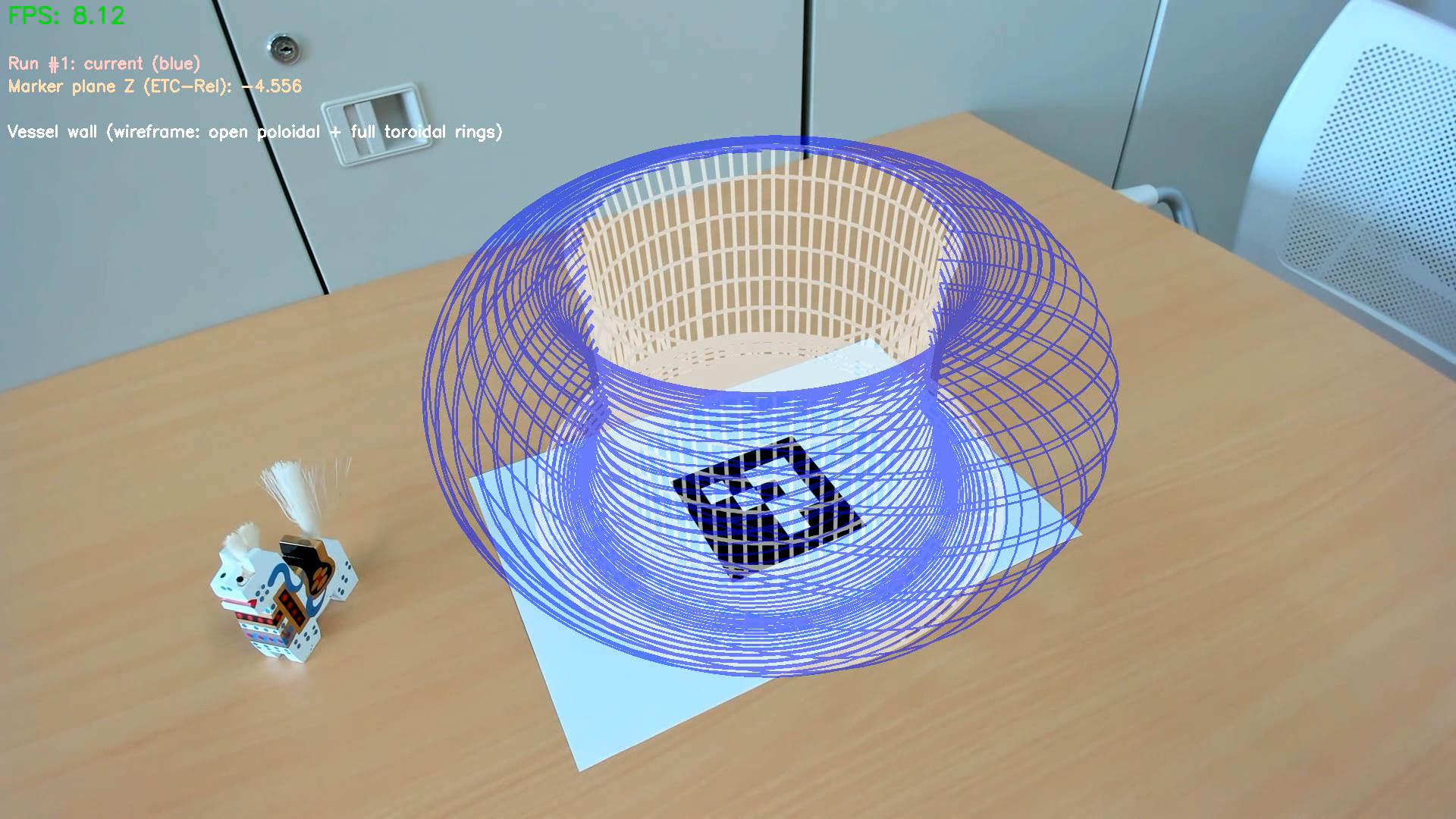}
    \caption{AR visualisation of the toroidal magnetic surface. Trajectories passing behind the inboard vessel wall are rendered with hidden-line processing.}
    \label{fig:Irrational_top}
  \end{center}
\end{figure}

\begin{figure}[htbp]
  \begin{center}
    \includegraphics[clip,width=14.0cm]{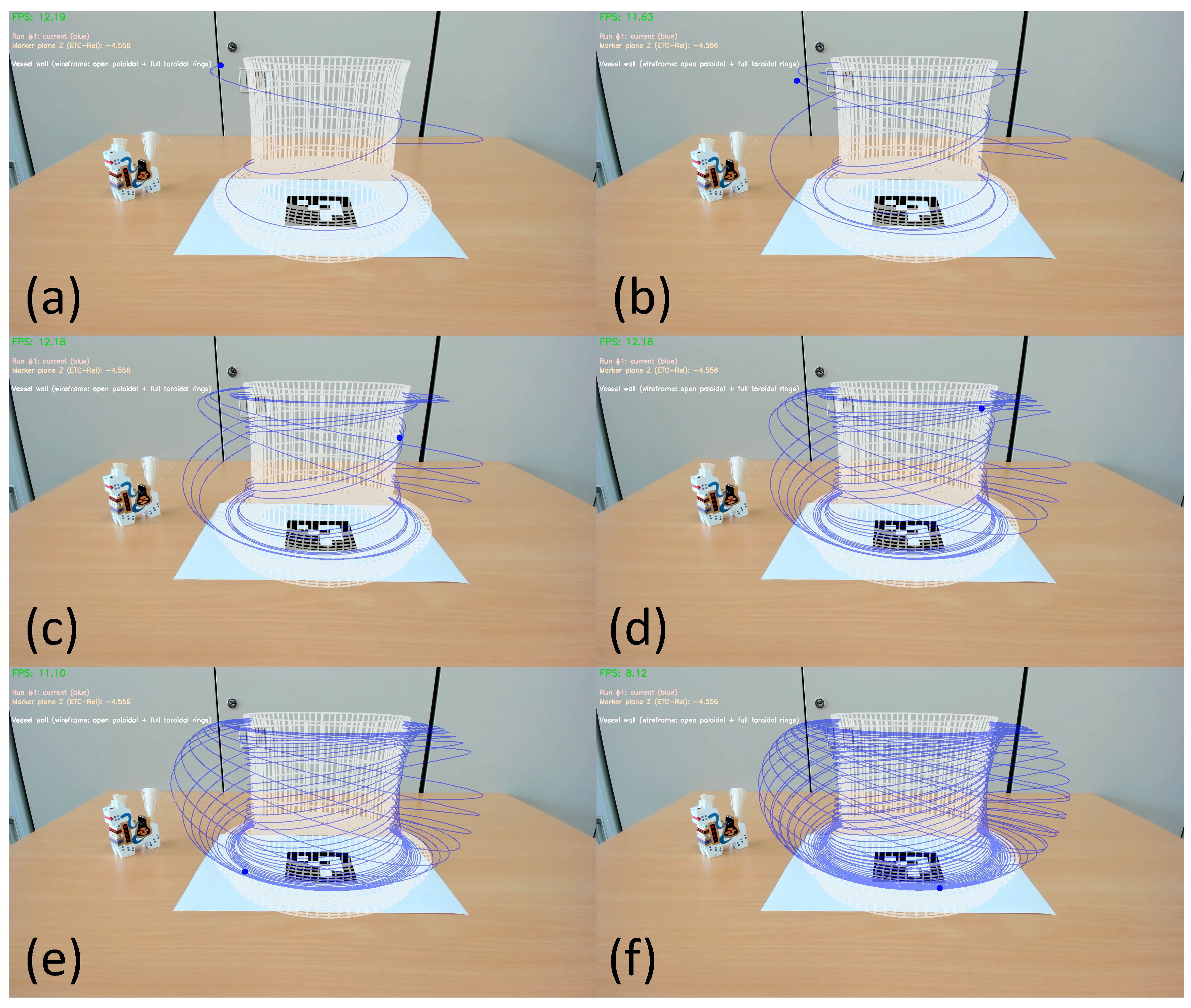}
    \caption{A sequence of visualisations of a single magnetic field line on an irrational magnetic surface. The FPS value indicates the effective frame rate of the complete processing and rendering loop. The decrease from 12.19 fps in (a) to 8.12 fps in (f) results mainly from the increasing rendering load as the accumulated trajectory becomes longer.}
    \label{fig:Irrational}
  \end{center}
\end{figure}

\begin{figure}[htbp]
  \begin{center}
    \includegraphics[clip,width=14.0cm]{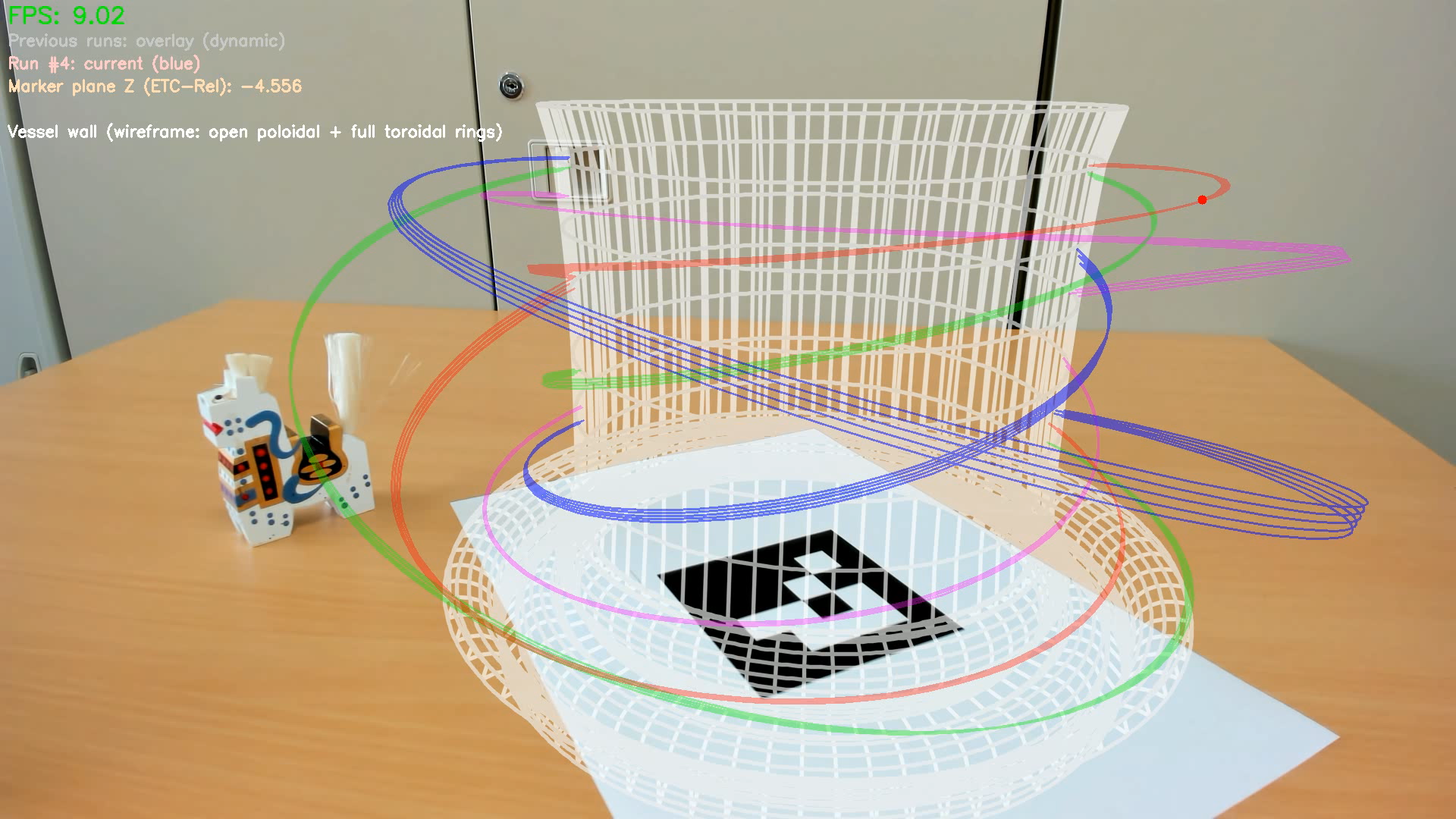}
    \caption{Visualisation of a rational magnetic surface with $\Delta \theta/\Delta \phi = 1/2$.}
    \label{fig:Rational}
  \end{center}
\end{figure}

\begin{figure}[htbp]
  \begin{center}
    \includegraphics[clip,width=14cm]{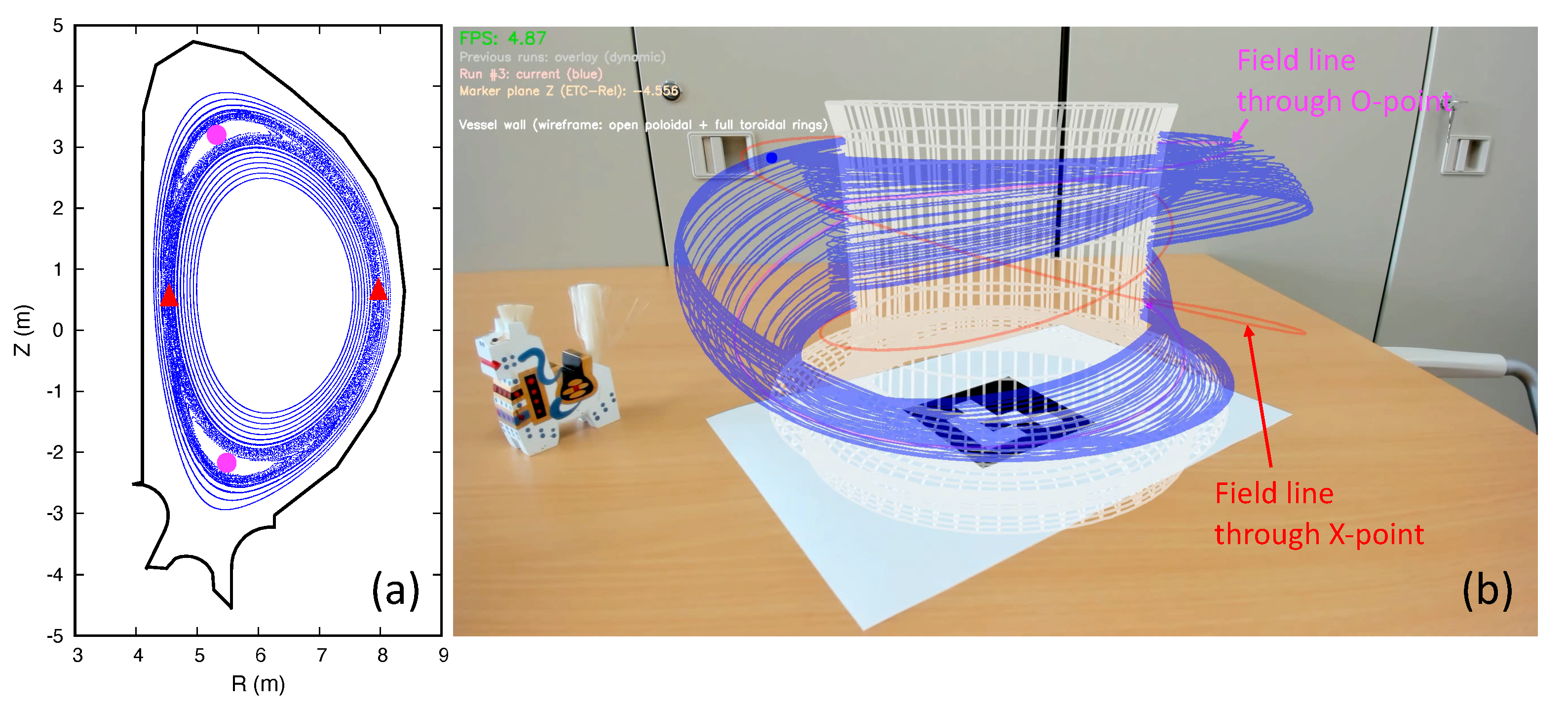}
    \caption{(a) Poincar\'{e} plot of a magnetic island. The approximate locations of the O- and X-points are indicated by a filled circle (magenta) and a filled triangle (red), respectively. (b) The corresponding AR visualisation. The magnetic field lines passing through the O- (magenta) and X-point (red) are also shown.}
    \label{fig:Island}
  \end{center}
\end{figure}

\begin{figure}[htbp]
  \begin{center}
    \includegraphics[clip,width=14cm]{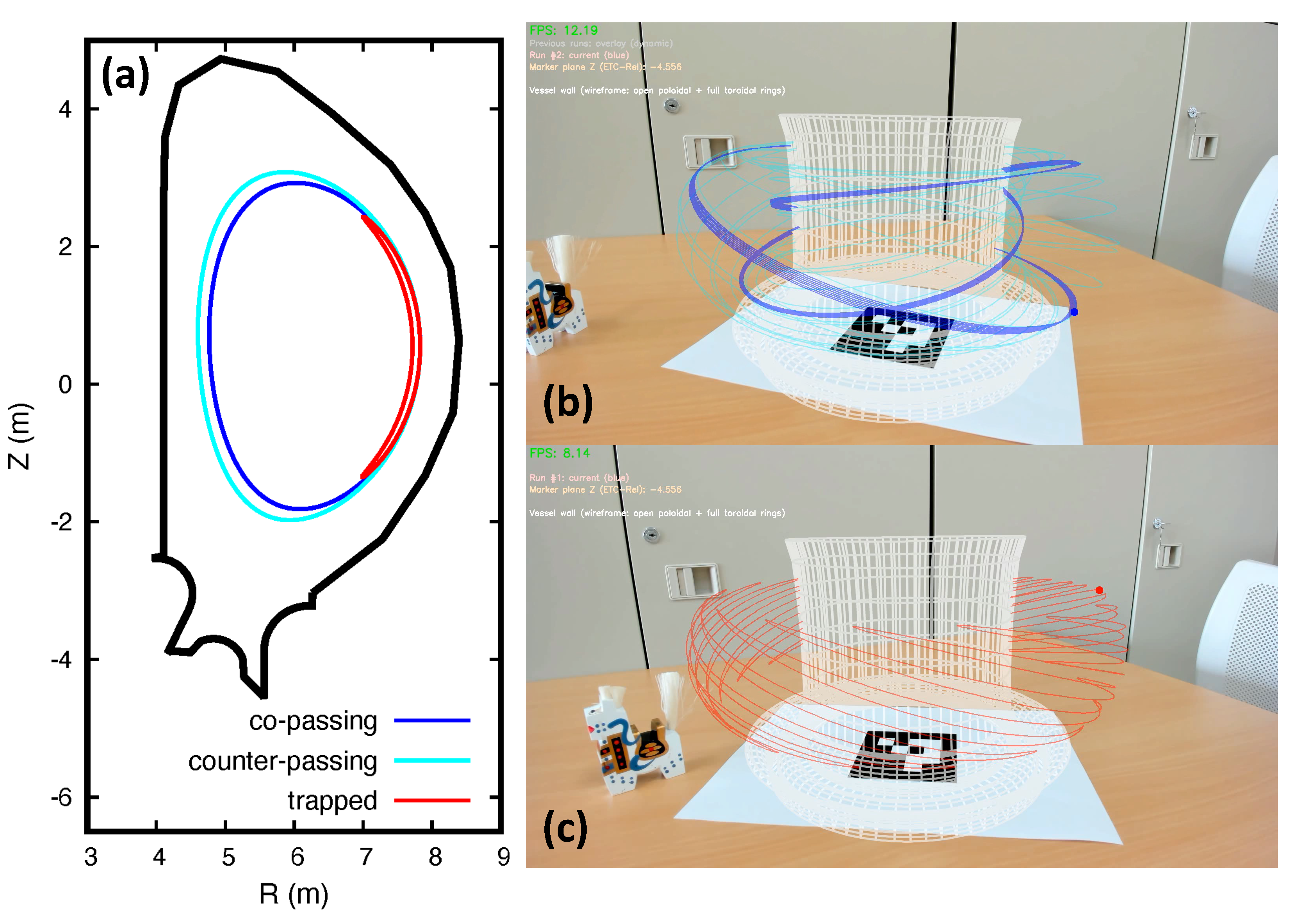}
    \caption{Charged-particle trajectories for a 3.5 MeV alpha particle. (a) Two-dimensional projection.  (b) Co-passing (${\bf v}\cdot {\bf B}>0$, moving along the magnetic-field direction) and counter-passing $({\bf v}\cdot {\bf B}<0$, moving opposite to it) particles. (c) Trapped (banana) particle trajectory.}
    \label{fig:ITER_AR}
  \end{center}
\end{figure}

From an educational perspective, the visualisations presented below can be associated with several specific learning goals. The visualisation of rational and irrational magnetic surfaces (Figs. \ref{fig:Irrational_top}--\ref{fig:Rational}) is intended to help students relate the two-dimensional representation of field-line intersections in a Poincar\'{e} plot to the corresponding three-dimensional trajectories and magnetic surfaces. The magnetic-island example (Fig. \ref{fig:Island}) extends this connection to the interpretation of O- and X-points (the centre and separatrix-crossing points of a magnetic island, respectively) and the three-dimensional structure of an island. The charged-particle examples (Fig. \ref{fig:ITER_AR}) are intended to connect poloidal projections of passing and trapped-particle orbits with their actual motion around the torus. In each case, the user can physically change the camera viewpoint while observing the same trajectory, allowing the structure to be examined from different directions rather than inferred from a single two-dimensional projection. The shared-display configuration also enables these visualisations to be used in instructor-led demonstrations or group activities, where an instructor and students can refer to and discuss the same visual information.

The AR representation is intended to complement, rather than replace, conventional two-dimensional representations such as Poincar\'{e} plots. Poincar\'{e} plots provide a compact and quantitative representation of field-line topology and are particularly effective for identifying fixed points, resonances, magnetic-island chains, and chaotic regions. In contrast, the AR representation provides direct access to the corresponding three-dimensional geometry and its temporal development as a field line is traced. Using the two representations together therefore allows the compact topological information provided by a Poincar\'{e} plot to be related to the three-dimensional spatial structure represented by the field-line trajectories.

Figure \ref{fig:Irrational_top} shows an example of AR visualisation of the magnetic field line trajectories. In an ideal tokamak configuration, there exists symmetry in the toroidal $(\zeta)$ direction, and the trajectories of the magnetic field lines in an equilibrium magnetic field are integrable. Therefore, when the trajectory is followed repeatedly, the corresponding Poincar\'{e} plot forms closed curves on the $\zeta=$ const. plane. These curves correspond to the toroidal surfaces in three dimensions. In this study, we adopt a method in which the inner side of the toroidal vessel wall and the lower structure of the device (called {\it divertor}) are drawn using meshes of different colours, and an occlusion rule is introduced such that trajectories that pass behind the vessel wall with respect to the major axis are not displayed. In this way, the characteristic features of the trajectories circulating around the torus can be recognised. Details of the occlusion rule are described in \ref{sec:Occlusion}.

It should be noted that the magnetic surface is represented here by the trajectory of a field line progressively tracing out the surface, rather than by an opaque rendered surface. Consequently, the front-back ordering of overlapping trajectories is not uniquely conveyed in a single static frame. In the present AR representation, this ambiguity is instead reduced by two dynamic cues: the temporal history of the traced field line and the change in its apparent geometry as the camera viewpoint is moved. These features provide additional cues for interpreting the three-dimensional arrangement, even though full surface-based occlusion is not implemented.

When the rotational transform is irrational, the magnetic field line gradually forms a surface structure as the number of toroidal turns increases. Figure \ref{fig:Irrational} shows the temporal evolution of the visualisation, in which a single magnetic field line is traced as it forms a closed magnetic surface near the plasma boundary. A higher density of magnetic field lines is observed on the inner side of the torus, which is related to the stability properties \cite{Biskamp} of the plasma. In contrast, when the starting point of the field-line tracing is chosen such that the rotational transform, namely the pitch $\Delta \theta/\Delta \phi$, is rational, the magnetic field line closes after a finite number of turns. Figure \ref{fig:Rational} presents a  visualisation example of such a rational surface, where magnetic field lines of different colours can be seen forming distinct closed trajectories. 

Figure \ref{fig:Island} illustrates an example of the visualisation of the magnetic island, comparing (a) a Poincar\'{e} plot and (b) the AR representation. As can be seen from the Poincar\'{e} plot [Figure \ref{fig:Island}(a)], isolated island structures appear around elliptic fixed points (O-points), whereas in the vicinity of hyperbolic fixed points (X-points), even small perturbations give rise to trajectories that diverge significantly from the X-point. This structure can be reduced to the behaviour of a one-degree-of-freedom Hamiltonian system, which is analogous to a simple pendulum \cite{Lichtenberg}. In the AR visualisation, when the initial position in the poloidal direction is varied at a radial location where the rotational transform takes a rational value, trajectories around both the O- and X-points are obtained. Figure \ref{fig:Island}(b) shows the trajectory of a magnetic field line traced from a point displaced from the X-point by $\Delta \theta/2\pi=0.10$; when the trajectory is followed for a sufficiently long time, a single field line forms an island structure that encloses the finite volume. 

Finally, we present the visualisation of the charged-particle orbit-following simulations. Figure \ref{fig:ITER_AR}(a) shows an example of a two-dimensional projection of the guiding-centre trajectories in a tokamak configuration. Trajectories of co-passing particles (${\bf v}\cdot {\bf B}>0$) and counter-passing particles (${\bf v}\cdot {\bf B}<0$) are plotted, and the corresponding AR representation is displayed in Fig. \ref{fig:ITER_AR}(b). Particles with sufficiently large velocity along the magnetic field circulate around the torus, but due to the inhomogeneity of the magnetic field, drift across the field occurs. As a result, depending on the sign of the velocity along the magnetic field, the trajectories deviate from the starting magnetic surface. Such deviations in trajectories are important for the control properties of magnetically confined fusion plasmas, such as the efficiency of plasma heating by energetic neutral beam injection \cite{Stix1972}. In the case of Fig. \ref{fig:ITER_AR}, the co-passing particle happens to close its orbit toroidally with a finite number of toroidal turns, but the counter-passing particle does not, which depends on the effective rotational transform experienced by passing particles with the guiding-centre drift. Figure \ref{fig:ITER_AR} (c) shows a trapped-particle, or “banana", orbit, in which the particle bounces between magnetic mirror points and forms a characteristic banana-shaped trajectory in the poloidal projection. The real-time feature of our AR system allows the user to recognise the actual guiding-centre velocities, which enables observing the time scale difference between the poloidal bounce motion and toroidal precession drift of the banana-particle centre. The example shown here has a relatively small orbit width, for which the bounce motion and toroidal precession can be clearly distinguished; the same visualisation method is also applicable to energetic-particle trajectories with larger finite orbit widths.

\section{Discussion and conclusion}\label{sec:DC}
In this study, we have presented a visualisation method that integrates an orbit-following simulation code with a marker-based AR system, and applied it to the analysis of magnetic field topology and charged-particle trajectories in magnetically confined fusion plasmas. The visualisation method introduced here possesses the characteristics of AR, namely, interactivity, real-time performance, and three-dimensionality. These features complement conventional visualisation methods using two-dimensional projections or Poincar\'{e} plots, where the users —especially, beginners— must infer three-dimensional structures from these two-dimensional representations. This mental reconstruction can impose additional cognitive load associated with spatial processing and representation \cite{Vidak2024}. The present AR environment enables the direct observation of three-dimensional spatial structures related to magnetic field lines and charged-particle trajectories while interactively manipulating the viewpoint. For magnetic field line tracing, the mapping property of the Poincar\'{e} map is replaced by a temporal evolution process, in which surfaces are formed sequentially around the magnetic surface, utilizing the real-time capability of AR. In the case of charged-particle visualisation, in addition to spatial trajectories, it is possible to perceive particles as moving objects that include velocity and temporal evolution.

In recent years, VR and AR technologies have also been increasingly explored for educational applications in physics \cite{Hammouch2026}. The system developed in this study can be implemented using only a web camera and display. Although similar visualisation can be realized using high-performance AR devices, such as head-mounted displays or smart glasses, the present approach can be regarded as cost-effective when considering applications in research, education, outreach, and citizen science. Display-based AR also allows multiple users to share the same visualisation results simultaneously, making it useful for collaborative group work. The specifications of the software and hardware used here represent only one example, and similar AR visualisation methods are expected to be applicable not only to magnetic confinement fusion but also to other fields of science and technology, such as accelerators, beam physics, and astrophysical objects. More generally, similar systems can be constructed for visualisation of streamlines in any types of three-dimensional vector field data.

The AR system has been used in practical exercises at the JT-60SA International Fusion School over three consecutive years. Each year, four students participated in a six-hour practical exercise conducted as four 1.5-hour sessions over two days. The students first performed camera calibration and then used the system to explore the examples described in this paper, including magnetic surfaces, rational and irrational surfaces, magnetic islands, and trapped-particle trajectories. In 2026, a short structured questionnaire was conducted after the practical exercise to obtain preliminary feedback from the participating students. 

The feedback provided several specific observations. For the charged-particle examples, students commented that the AR representation helped them recognise the three-dimensional motion of trapped particles, including their back-and-forth motion. For the magnetic-field examples, some students reported that changing the viewpoint helped them relate magnetic surfaces and islands to their corresponding Poincar\'{e} plots. One student also reported that, after the AR exercise, two-dimensional plots were interpreted by mentally reconstructing the corresponding three-dimensional particle trajectories. The shared display was also reported to support discussion among the students. Importantly, the feedback also indicated limitations of the AR representation: one student found magnetic islands easier to interpret quantitatively in a conventional two-dimensional plot. In response to the questionnaire item ``Was there anything that remained difficult to understand in AR, or that was actually easier to understand using a conventional 2D plot?'', one student commented, ``In the Poincare plot, I could see the structures of many magnetic field lines on the poloidal cross-section at the same time. In AR, on the other hand, I could understand their three-dimensional structures more clearly, but I could see fewer field lines at once.'' Another student wrote, ``Magnetic islands are hard to see using 3d plot, even from different angles, 2D remains better in that case.'' These observations suggest that the AR representation may be most useful as a complement to, rather than a replacement for, conventional two-dimensional visualisations. Given the small number of participants and the absence of a controlled pre/post assessment, this feedback should be regarded as exploratory and does not establish educational effectiveness.

Regarding technical aspects, it would be worth noting that present AR visualisation involves trade-offs among the amount of information, fidelity of the visualised structure (such as occlusion), representation methods (point clouds, polygons, volume data), and device constraints (cost and size). In this study, a simple occlusion rule based on the fact that magnetic field lines and charged particles circulate around a toroidal device was introduced. However, the present implementation uses trajectory-based two-dimensional drawing rather than surface-based three-dimensional rendering and does not utilize the depth information of the camera image, making it difficult to implement complex occlusion processing. Front-back ordering can therefore remain ambiguous in a static view, while the temporal evolution of the trajectories and interactive changes in viewpoint provide additional cues for interpreting their three-dimensional arrangement. Accordingly, the present system is primarily suited for qualitative exploration of three-dimensional spatial structures, rather than for precise quantitative analysis of complex nested surfaces or their front-back relationships. Quantitative analysis should therefore continue to rely on conventional representations and analysis tools, with the AR visualisation serving as a complementary means of exploring their three-dimensional spatial interpretation. Further improvements are required to expand the range of visualisation capability.

\appendix
\section{Occlusion law}\label{sec:Occlusion}
Let the camera position be ${\bf C}=(C_x,C_y,C_z)$, and a point on the trajectory to be observed be ${\bf P}=(P_x,P_y,P_z)$. A point on the ray connecting them can be written using the velocity vector ${\bf v}={\bf P}-{\bf C}$ as
\begin{equation}
 {\bf X}(\lambda) ={\bf C} + \lambda {\bf v}, \quad (0\leq \lambda \leq 1).
\end{equation}
The inboard surface of the toroidal vessel of radius $R_p$ is given by $x^2+y^2=R_p^2$, satisfying $z_{\rm min}\leq z \leq z_{\rm max}$ in the vertical direction. Whether the ray intersects the side surface can be determined from the quadratic equation
\begin{equation}
 (v_x^2+v_y^2)\lambda^2 + 2(C_xv_x+ C_y v_y) \lambda + (C_x^2+C_y^2-R_p^2) = 0.
\end{equation}
If the discriminant is non-negative, and the solution $\lambda$ satisfies $0<\lambda <1$ and 
\begin{equation}
 z_{\rm min} -\epsilon \leq C_z +\lambda v_z \leq z_{\rm max}+\epsilon,
\end{equation}
then the intersection point lies in front of the camera and in front of the point ${\bf P}$. Therefore, ${\bf P}$ is judged to be occluded by the vessel wall. Here, $\epsilon= 0.05$ m is a small margin applied to the axial extent of the cylinder in the visibility judgement. The present rule treats only the inboard cylindrical wall and does not account for occlusion by the complete three-dimensional vessel geometry.

\ack
The author would like to thank Hirotake Ishii for his advice on extended reality techniques. The author is also grateful to the organisers and secretaries of the JT-60SA International Fusion School (JIFS) for providing the opportunity to conduct the AR practical exercise over three consecutive years, and in particular to Eva Belonohy for her continuous encouragement. The author would also like to thank all twelve students who participated in the practical exercises during these three years. In particular, Piotr Pietrzak, Rajbir Kaur, Yuta Kinashi, and Tomomi Nagai, who participated in the 2026 exercise, are thanked for providing valuable feedback on the AR visualisation system, which helped improve the manuscript. The author also acknowledges Louis Puel for his contribution to the development of the visualisation concept illustrated in Fig. \ref{fig:Island}. 

The JT-60SA International Fusion School (JIFS) has been carried out in collaboration between QST and the EUROfusion Consortium, funded by the European Union via the Euratom Research and Training Programme (Grant Agreement No 101052200 — EUROfusion). Views and opinions expressed are however those of the author(s) only and do not necessarily reflect those of the European Union or the European Commission. Neither the European Union nor the European Commission can be held responsible for them. This work was supported in part by the NIFS Collaboration Research Program (NIFS24KUHL131) and by Grants-in-Aid for Scientific Research (JSPS KAKENHI Grant No. 25K00982).

\end{document}